# Quantum mechanical aspects of friction and electric resistance in microscopic problems with applications to radiation physics

W. Ulmer

Strahlentherapie Nordwürttemberg, Germany, and Research Institute for Medical and MPI of Physics, Göttingen, Germany

**Abstract**

Friction incorporates the close connection between classical mechanics in irreversible thermodynamics. The translation to a quantum mechanical foundation is not trivial and requires a generalization of the Lagrange function. A change to electromagnetic circuits appears to more adequate, since the electric analogue (Ohm's law) is related to scatter of electrons at latte vibrations.

1. Introduction

With regard to the motion of macroscopic systems friction plays a significant role in all dynamical processes of classical mechanics, since friction is connected to energy dissipation, which may be considered as a special case of the Langevin equation [1 - 3]:

$$M \cdot \ddot{q} + \gamma \cdot \dot{q} + \frac{\partial U}{\partial q} = 0 \quad (1.1)$$

It is usual to start with the Lagrange formalism and to write eq. (1.1) in the following way:

$$Ł_0 = \frac{M}{2} \cdot \dot{q}^2 - U(q)$$
$$\frac{d}{dt} \cdot \frac{\partial Ł_0}{\partial \dot{q}} - \frac{\partial Ł_0}{\partial q} = D(\gamma, \dot{q}) \quad (1.2)$$

$Ł_0$ represents the 'standard' Lagrangian without friction, D represents a dissipation function, and the motion of a free mass M (U = 0) suffers continuously slowing down with the range R given by (boundary conditions at t = 0 are q = 0 and $\dot{q} = v_0$):

$$q(t) = R - (v_0 \cdot M / \gamma) \cdot exp(-\gamma \cdot t / M)$$
$$q(t \to \infty): \quad R = \frac{v_0 \cdot M}{\gamma} \quad (1.3)$$

In the past decades, many attempts have been put forward to describe friction properties by a nonlinear Schrödinger equation with a logarithmic nonlinearity [4 - 7]:

$$i \cdot \hbar \frac{\partial}{\partial t} \psi - H_0 \psi = \lambda \cdot [i \cdot ln(\psi/\psi^*) \cdot \psi + i \cdot \langle ln(\psi/\psi^*) \rangle] \psi \quad (1.4)$$

$H_0$ refers to the standard Hamiltonian, which appears in the linear Schrödinger equation. This extension leads to the problem, in which way this equation is applicable in the microscopic domain. Thus the stopping of fast particles, e.g. protons or electrons can be quantum mechanically described by the Bethe-Bloch equation, which reads:

$$-\frac{dE(z)}{dz} = \frac{K}{v^2} \cdot [\ln(\frac{2mv^2}{E_I}) - \ln(1-\beta^2) + cor\_terms]$$

$$K = \frac{Z \cdot \rho}{2 \cdot A_N \cdot m} \cdot 8\pi \cdot e_0^4 \cdot q_p^2 \qquad (1.5)$$

$$cor\_terms : a_{shell} + a_{Barkas} + a_{Bloch} \qquad (1.5a)$$

All terms of eq. (1.5) and eq. (1.5a) inclusive the integration-procedure has been presented in detail [8 - 9]. The contributions $a_{shell}$, $a_{Barkas}$ and $a_{Bloch}$ represent correction terms; ρ is the density of the medium (e.g. water (ρ =1 g/cm$^3$) is often a reference medium), Z and $A_N$ are the effective charge/mass number of the medium, $q_p$ is the charge of the projectile particle, $e_0$ the elementary charge, $E_I$ an averaged ionization energy by inclusion of all possible energy levels of the medium and ß = v/c. All calculations of *E(z)* of projectile particles are based on eq. (1.5). It appears that the conception of friction for slowing down of microscopic particles can hardly be included by the classical friction. It should be noted that a modified version of eq. (1.1) with a completely different friction term can be used to derive empirical properties of slowing down of the motion of protons and neutrons:

$$M \cdot \ddot{q} + \delta \cdot \dot{q}^{-\beta} = 0 \qquad (1.6)$$

The integration of this equation is straightforward and leads to the range formula (<u>c</u>ontinuous <u>s</u>lowing <u>d</u>own <u>a</u>pproximation → csda):

$$R_{csda} = A \cdot E_0^p \qquad (1.7)$$

A is a constant factor (see eq. 1.7)), $E_0$ is the initial energy and p ≈1.74 for protons and p = 1.5 (ß = 1) for slow neutrons (Geiger rule). The empirical energy – range relation can be derived by the above formula. With regard to fast electrons ($E_{kinetic}$ > mc$^2$) and fast protons ($E_{kinetic}$ > 200 MeV) formula (1.6) has to be modified to include relativistic corrections (a derivation of the relativistic modification of eq. (1.5) is given in [8]):

$$R_{csda} = A \cdot (E_0 + E_0^2 / 2Mc^2)^p$$

$$p = 1 + \beta/2; \quad A = 2^{p-1} / (\delta \cdot M^{p-1}) \qquad (1.8)$$

Eqs. (1.5 – 1.7) show that fast micro-particles are stopped by passing through a medium (e.g. water) with an energy loss being proportional to a power of the inverse velocity, i.e. ≈ 1/v$^\beta$ with 1≤ ß < 2. The fact is, in particular, true for the Bethe-Bloch eq. (1.5). However, this equation indicates that the stopping power - *dE(z)/dz* cannot be simply reduced to one power of the velocity *v* or the actual kinetic energy *E(z)*, since the correction terms denoted by '*cor_terms*' lead to intricate modifications of the stopping power. Thus a very accurate adaptation of the solutions given by eq. (1.5) via eq. (1.8) indicates that the power *p* is slightly depending on the initial energy $E_0$ of the projectile particle, which may vary between *p = 1.70* and *p = 1.76* [7]. It has also to be mentioned that eqs. (1.5 – 1.8) are restricted to '*csda*'. In order to determine the energy loss - *dE/dz* of a projectile particle by passing through a medium energy straggling and lateral scatter have to be included [8], where further details can be obtained.

Therefore the question arises, whether eq. (1.1) is at all adequate in the molecular level. We mainly think of molecular oscillation with a reduced mass in the normal mode. Usually the energy of these oscillations can be restricted to the domain of thermal energy (≈ $k_B \cdot T$), and then the friction term of classical physics may be applicable. This motion can be damped by collisions with neighboring molecules or energy transfer via dipole-dipole interactions (van der Waals interactions). Classically this equation of motions reads:

$$M \cdot \ddot{q} + \gamma \cdot \dot{q} + M \cdot \omega_0^2 \cdot q = 0 \quad \text{with} \quad \omega_0^2 = f/M \qquad (1.8)$$

Thus *f* represents the force constant, and the reduced frequency of eq. (1.8) is given by:

$$\omega^2 = \omega_0^2 - \gamma^2 / (4 \cdot M^2) \qquad (1.9)$$

This equation makes also sense in quantum mechanics; it requires a modification of the Lagrange function $L_0$ we consider in the following section. It should be mentioned that the consideration of friction (Ohm's law) in electromagnetic circuits appears to be more adequate for a quantum theoretical treatment.

## 2. Modification of the Lagrange function $Ł_0$ to yield the Lagrange-Hamilton formalism or the Schrödinger equation with irreversible correction terms.

The requested generalization of the Lagrange function is given by:

$$Ł = Ł_0 \cdot exp(\gamma \cdot t / M) \quad (2.1)$$

Thus the application of the Lagrange formalism according to eq. (1.2) with regard to eq. (2.1) yields eq. (1.1). The determination of the canonical momentum $P$ due to eq. (2.1) and using the Hamiltonian formalism yields:

$$H = exp(-\gamma \cdot t / M) \cdot P^2 / 2M + V(q) \cdot exp(\gamma \cdot t / M) \quad (2.2)$$

The Schrödinger equation of the Hamiltonian (2.2) reads:

$$-\frac{\hbar^2}{2M} \cdot exp(-\gamma \cdot t / M) \frac{\partial^2}{\partial q^2} \psi(q,t) + V(q) \cdot exp(\gamma \cdot t / M) \cdot \psi(q,t) = i \cdot \hbar \cdot \frac{\partial}{\partial t} \cdot \psi(q,t) \quad (2.3)$$

Although the physical sense may be questionable, we consider first the solution of a free particle with V(q) = 0. The particle mass M should be much bigger than that of a proton mass and the velocity v << c. Thus the motion of molecules in a medium appears to be adequate; the phenomenological friction term $\gamma$ may be realized by interactions with the environment (e.g. a solvent).

In a first step, we start with the *'ansatz'* of a 'free' particle:

$$\psi(q,t) = A(t) \cdot exp(-i \cdot k \cdot q) \quad (2.4)$$

Inserting eq. (2.4) into eq. (2.3) provides:

$$\psi(q,t) = A_0 \cdot exp(-i \cdot k \cdot q) \cdot exp[-\frac{i \cdot \hbar \cdot k^2 (1 - exp(-\gamma \cdot t / M))}{2\gamma}] \quad (2.5)$$

However, this solution (2.5) has the disadvantage that for t →∞ a comparison with eq. (1.3) is hardly possible, whereas the low order expansion of (1- exp(-$\gamma \cdot$t/M))/2$\gamma$) makes sense:

$$exp[-\frac{i \cdot \hbar \cdot k^2 (1 - exp(-\gamma \cdot t / M))}{2\gamma}] \approx exp(-\frac{i \cdot \hbar \cdot k^2 \cdot t}{2M}) \quad (2.6)$$

Eq. (2.6) incorporates the initial behavior at rather small time intervals. In order to be free of this restriction, the integration over all possible '*k-values*' has to be performed:

$$\varepsilon = \sqrt{1 - exp(-\gamma \cdot t / M)} \quad (2.7)$$

$$\psi(q,t) = \frac{A_0}{2 \cdot \varepsilon \cdot \sqrt{\pi}} \cdot \sqrt{\frac{2\gamma}{i \cdot \hbar}} exp(\frac{i \cdot \gamma \cdot q'^2}{2\hbar \cdot \varepsilon^2})$$

$$\text{where}: q' = q - \bar{q} \quad (2.8)$$

The normalization of the wave-function yields:

$$A_0^2 = \frac{2\pi \varepsilon^2 \hbar}{\gamma \cdot \bar{q}} \quad (2.9)$$

The range relation (1.3) results by forming the expectation value <p>/M at t → ∞.

A typical task is the damped harmonic oscillator $V(q) = M \cdot \omega_0^2 \cdot q^2/2$. This type of potential is similar to the motion of charged molecules in constant magnetic fields. By that, we obtain the quantum mechanical analogue of eqs. (1.8 and 1.9). The Schrödinger equation of the damped harmonic oscillator reads:

$$-\frac{\hbar^2}{2M} \cdot exp(-\gamma \cdot t / M) \cdot \frac{\partial^2}{\partial q^2} \psi(q,t) + \frac{M \cdot \omega^2}{2} \cdot q^2 \cdot exp(\gamma \cdot t / M) \cdot \psi(q,t) = i\hbar \frac{\partial}{\partial t} \psi(q,t) \quad (2.10)$$

This problem is identical to the charge quantization problem with a damping term (Ohm's). The ground state properties of the damped harmonic oscillator are given by:

$$\psi_0(q,t) = \exp[-\tfrac{1}{2}\xi^2] \cdot \exp(i \cdot \tfrac{1}{2} \cdot a \cdot t) \quad (2.11)$$

$$\xi^2 = \tfrac{M \cdot \omega}{\hbar} \cdot q^2 \cdot \exp(\gamma \cdot t / M); \quad \omega^2 = \omega_0^2 - \tfrac{\gamma^2}{4 \cdot M^2}; \quad a = [\tfrac{i \cdot \gamma}{2M} \pm \omega] \quad (2.12)$$

Thus the general solution can readily be constructed from the ground state by introducing Hermite polynomials:

$$\psi_n(q,t) = H_n(\xi) \cdot \exp(-\tfrac{1}{2} \cdot \xi^2) \cdot \exp(i \cdot (n+1/2) \cdot a \cdot t) \quad (2.13)$$

Eqs. (2.11 - 2.13) possess the same reduced frequency ω as in the classical case (1.9), and due to friction stationary states do not exist. The behavior of these solutions can be characterized by switching on the friction at the initial condition t = 0. It should be mentioned that with regard to different solutions $\psi_n$ and $\psi^*_m$ the orthogonal relations exist. An important feature is the motion of a charged particle (e.g. an ionic molecule) in a constant magnetic with an additional damping. The Hamiltonian of this problem reads:

$$\tfrac{1}{2M} \exp(-\gamma \cdot t / M) \cdot (\tfrac{\hbar}{i} \nabla - \tfrac{q_0}{c} \vec{A})^2 \psi = i\hbar \tfrac{\partial}{\partial t} \psi \quad (2.14)$$

Denoting the coordinates by *(x, y, z)* and $q_0$ the electric charge of an ion, we define the vector potential $\vec{A}$ by:

$$A_x = -B_0 \cdot y; \quad B_x = B_y = 0; \quad B_z = -\tfrac{\partial}{\partial y} A_x = B_0 \quad (2.15)$$

By that, the above eq. (2.14) assumes the shape:

$$\exp(-\gamma \cdot t / M) \cdot [-\tfrac{\hbar^2}{2M} \Delta - \tfrac{i \cdot \hbar \cdot q_0}{M \cdot c} B_0 \cdot y \cdot \tfrac{\partial}{\partial x} + \tfrac{q_0^2}{2Mc^2} B_0^2 \cdot y^2] \psi = i\hbar \tfrac{\partial}{\partial t} \psi \quad (2.16)$$

The solution of eq. (2.16) can be found in every textbook of quantum mechanics, if the friction term γ is 0 (stationary case), and the Larmor frequency would assume the form:

$$\omega_0 = \tfrac{q_0 \cdot B_0}{M \cdot c} \quad (2.17)$$

By taking account of γ ≠ 0 we have to modify the 'standard' solution by:

$$\psi(x, y, z, t) = A(t) \cdot \exp(i \cdot (\alpha \cdot x + \beta \cdot z)) \cdot \varphi(y', t) \quad (2.18)$$

$$y' = y - \tfrac{\hbar \cdot \alpha \cdot c}{q_0 \cdot B_0} \quad (2.19)$$

This substitution provides the well-known oscillator equation with inclusion of friction:

$$\exp(-\gamma \cdot t / M) \cdot [-\tfrac{\hbar^2}{2M} \tfrac{\partial^2}{\partial y'^2} + \tfrac{M \cdot \omega^2}{2} \cdot y'^2] \varphi = i\hbar \tfrac{\partial}{\partial t} \varphi \quad (2.20)$$

This equation can be solved by the previously elaborated methods to yield:

$$\varphi_n(\xi, t) = A_0 \cdot H_n(\xi) \cdot \exp(-\tfrac{\xi^2}{2}) \cdot \exp[\tfrac{M}{i \cdot \hbar \cdot \gamma} \cdot (1 - \exp(-\tfrac{\gamma \cdot t}{M}))]$$

$$\omega^2 = \omega_0^2 - \tfrac{\gamma^2}{4M^2}; \quad \xi^2 = \tfrac{M \cdot \omega}{\hbar} \cdot y'^2 = \tfrac{M \cdot \omega}{\hbar} \cdot (y + \tfrac{\hbar \cdot \alpha \cdot c}{q_0 \cdot B_0})^2 \quad (2.21)$$

The terms referring to exp(i·α·x) and exp(i·β·z) have already been given by the previous methods (i.e. *motion of free particles with friction)* and can readily be added to obtain the complete wave-function ψ. By the substitutions *M → L*, $\omega_0^2 \to 1/LC$, q(position) *→Q(electric charge)*, *γ(friction)* → R(Ohm's resistance) the treatment of damped circuits is straightforward.

### 3. Quantization of electromagnetic circuits with damping (Ohm's law)

The extension of Lagrange formalism to electromagnetic circuits is more promising. A main reason comes from molecular physics/biophysics, since the charge distribution in a molecule (or molecular environment) can be treated as a capacity $C$, whereas transitions to different states represent electric currents associated with the inductivity $L$. Very impressive examples are the H bonds between the DNA base pairs, where the exchange protons represent a tunnel current and simultaneously induce a weak local magnetic field [10, 11].

At first, we regard Figure 1A with the case $R \neq 0$. Thus the generalized Lagrange function $Ł$ is given by:

$$Ł = Ł_0 \cdot exp(R \cdot t / L)$$
$$Ł_0 = \tfrac{L}{2} \cdot \dot{Q}^2 - \tfrac{L \cdot \omega_0^2}{2} \cdot Q^2; \quad \omega_0^2 = 1/(LC) \quad (3.1)$$

The Hamilton operator and the Schrödinger equation in the charge space are obtained with the help of the canonical momentum:

$$P = \tfrac{\partial}{\partial \dot{Q}} \cdot Ł(Q, \dot{Q}, t) \quad (3.2)$$

It should be pointed out that the canonical momentum P is equivalent to the magnetic flux Φ.

Thus the time-dependent Schrödinger now assumes the shape:

$$-\tfrac{\hbar^2}{2 \cdot L} \cdot exp(-R \cdot t / L) \cdot \tfrac{\partial^2}{\partial Q^2} \cdot \psi(Q,t) + \tfrac{L \cdot \omega_0^2}{2} \cdot Q^2 \cdot exp(R \cdot t / L) \cdot \psi(Q,t) = i \cdot \hbar \tfrac{\partial}{\partial t} \cdot \psi(Q,t) \quad (3.3)$$

According to previous results the general solution is:

$$\psi_n(Q,t) = H_n(\xi) \cdot exp(-\tfrac{1}{2} \cdot \xi^2) \cdot exp(i \cdot (n+1/2) \cdot a \cdot t) \quad (n=0,1,,,,.);$$
$$\xi^2 = \tfrac{L \cdot \omega}{\hbar} \cdot Q^2 \cdot exp(R \cdot t / L); \quad \omega_0^2 = \tfrac{1}{LC}; \quad \omega^2 = \omega_0^2 - \tfrac{R^2}{4 \cdot L^2}; \quad a = [\tfrac{i \cdot R}{2L} \pm \omega] \quad (3.4)$$

The case with two magnetically coupled electric circuits (the mutual inductivity between the circuits is denoted by $M$ the electric charge by $Q$, see Figure 1B). It should also be pointed out that the electric resistance (Ohm's law) and the related heat production can be regarded as an obvious microscopic problem, since it results from collisions of electrons with lattice vibrations. Therefore the quantum mechanical consideration is justified.

In absence of an electric resistance (R = 0) the equations for the two coupled circuits are given by:

$$L \cdot \ddot{Q}_1 + M \cdot \ddot{Q}_2 + Q_1/C = 0$$
$$L \cdot \ddot{Q}_2 + M \cdot \ddot{Q}_1 + Q_2/C = 0 \quad (3.5)$$

The normal modes are given by the substitutions:

$$q_1 = Q_1 - Q_2; \quad q_2 = Q_1 + Q_2 \quad (3.6)$$

The Lagrange function of the two resulting equations for the normal modes is given by:

$$Ł_1 = \tfrac{\lambda_1}{2} \cdot \dot{q}_1^2 - \tfrac{\lambda_1 \cdot \omega_1^2}{2} q_1^2; \quad Ł_2 = \tfrac{\lambda_2}{2} \cdot \dot{q}_2^2 - \tfrac{\lambda_2 \cdot \omega_2^2}{2} q_2^2$$
$$\lambda_1 = L - M; \quad \lambda_2 = L + M; \quad \omega_{1,2}^2 = \tfrac{1}{C \cdot \lambda_{1,2}} \quad (3.7)$$

Thus the Lagrange functions $Ł_1$ and $Ł_2$ provide the tools to determine the canonical momentum $p_1$ and $p_2$ and the application of canonical commutation relations: $[p_1, q_1] = [p_2, q_2] = -i \cdot \hbar$. At this place, we do not consider the related Schrödinger equation in the charge space. However, with the help of creation – annihilation operators the problem can be treated as known from the oscillators in the position space:

$$b_k = \alpha_k \cdot q_k + i\beta_k p_k$$
$$b_k^+ = \alpha_k \cdot q_k - i\beta_k p_k$$
$$\alpha_k^2 = \frac{\lambda_k \cdot \omega_k}{2 \cdot \hbar} \; ; \; \beta_k^2 = \frac{1}{2 \cdot \hbar \cdot \lambda_k \cdot \omega_k} ; \; (k=1,2) \quad (3.8)$$

The commutation relations of the operators $b_k$ and $b_k^+$ are well-known, i.e. $[b_k, b_k^+] = 1$ (k = 1, 2), and the Hamilton operator of the normal modes reads:

$$H = \hbar \cdot \omega_1 \cdot (n+1/2) \cdot b_1^+ \cdot b_1 + \hbar \cdot \omega_2 \cdot (m+1/2) \cdot b_2^+ \cdot b_2$$
$$(n,m = 0,1,2...) \quad (3.9)$$

In the presence of the Ohm's resistance with $R \neq 0$ we may either solve the Schrödinger equation for each mode (k = 1, 2 according to eqs. (2.11, 2.12)) or modify the creation – and annihilation operators. The Hamiltonian now becomes:

$$H = [\sum_{k=1}^{2} \frac{1}{2 \cdot \lambda_k} \cdot p_k^2 \cdot \exp(-\frac{R \cdot t}{\lambda_k}) + \frac{\lambda_k \cdot \omega_k^2}{2} \cdot q_k^2 \cdot \exp(\frac{R \cdot t}{\lambda_k})] \quad (3.10)$$

In order to formulate eq. (3.10) according to eq. (3.8) we have to perform the following modification:

$$b_k'(t) = \alpha_k \cdot q_k \cdot \exp(\frac{R \cdot t}{2 \cdot \lambda_k}) + i\beta_k p_k \cdot \exp(-\frac{R \cdot t}{2 \cdot \lambda_k});$$
$$b_k^{+\prime}(t) = \alpha_k \cdot q_k \cdot \exp(\frac{R \cdot t}{2 \cdot \lambda_k}) - i\beta_k p_k \cdot \exp(-\frac{R \cdot t}{2 \cdot \lambda_k});$$
$$b_k'(t) b_k^{+\prime}(t) - b_k^{+\prime}(t) b_k'(t) = 1 \quad (k=1,2) \quad (3.11)$$
$$\omega_k^2 = \frac{1}{C \cdot \lambda_k} - \frac{R^2}{4 \cdot \lambda_k^2} \quad (k=1,2) \quad (3.12)$$

However, a stationary form like eq. (3.9) does not exist, and therefore we have to write the problem similar to the time-dependent Schrödinger equation:

$$H \cdot \psi = \frac{\hbar}{2} \sum_{k=1}^{2} \omega_k \cdot [b_k'(t) b_k^{+\prime}(t) + b_k^{+\prime}(t) b_k'(t)] \cdot \psi = i\hbar \frac{\partial}{\partial t} \psi \quad (3.13)$$

Thus the handling of eq. (3.13) according to the algebra (3.11) is a little more difficult to calculate excited states and transitions, since the time-dependent version has to be accounted for each excited state $\psi_n(b_k'(t), b_k^{+\prime}(t), t)$. The procedure is equivalent to that of eq. (3.4).

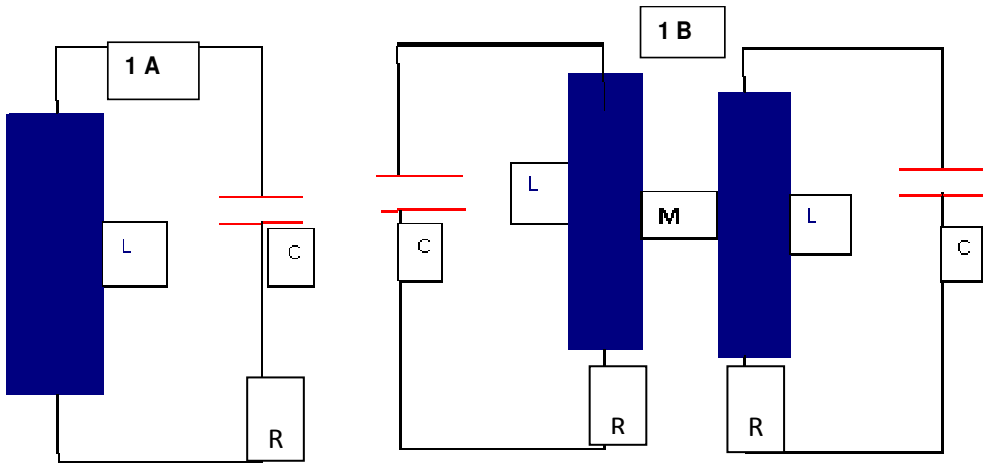

**Figure 1:** **A.** One single oscillator (L: inductivity and C: capacity). B. Two identical oscillators with mutual coupling M of the currents (magnetic coupling). R: Ohm's resistance.

## 4. Coupled oscillator circuits according to Figure 2 and Figure 3 and a damped oscillator forced by the voltage $V_f$

Figure 2 presents three identical oscillators, which are mutually coupled by the interaction inductivity M. The well-known dynamical equations are:

$L \cdot \ddot{Q}_1 + M \cdot (\ddot{Q}_2 + \ddot{Q}_3) + Q_1/C = 0$ (4.1)

$L \cdot \ddot{Q}_2 + M \cdot (\ddot{Q}_1 + \ddot{Q}_3) + Q_2/C = 0$ (4.2)

$L \cdot \ddot{Q}_3 + M \cdot (\ddot{Q}_1 + \ddot{Q}_2) + Q_3/C = 0$ (4.3)

With the help of the following substitutions we obtain the normal modes:

$q_1 = Q_1 - Q_3$; $q_2 = Q_2 - Q_1$; $q_3 = Q_1 + Q_2 + Q_3$ (4.4)

These substitutions imply the Lagrange function of the normal modes:

$Ł_1 = \frac{1}{2}\lambda_1 \cdot \dot{q}_1^2 - \frac{\lambda_1}{2}\omega_1^2 \cdot q_1^2$; ($\lambda_1 = L - M$; $\omega_1^2 = \frac{1}{\lambda_1 \cdot C}$) (4.5)

$Ł_2 = \frac{1}{2}\lambda_2 \cdot \dot{q}_2^2 - \frac{\lambda_2}{2}\omega_2^2 \cdot q_2^2$; ($\lambda_2 = L - M$; $\omega_2^2 = \frac{1}{\lambda_2 \cdot C}$) (4.6)

$Ł_3 = \frac{1}{2}\lambda_3 \cdot \dot{q}_3^2 - \frac{\lambda_3}{2}\omega_3^2 \cdot q_3^2$; ($\lambda_3 = L + 2 \cdot M$; $\omega_3^2 = \frac{1}{\lambda_3 \cdot C}$) (4.7)

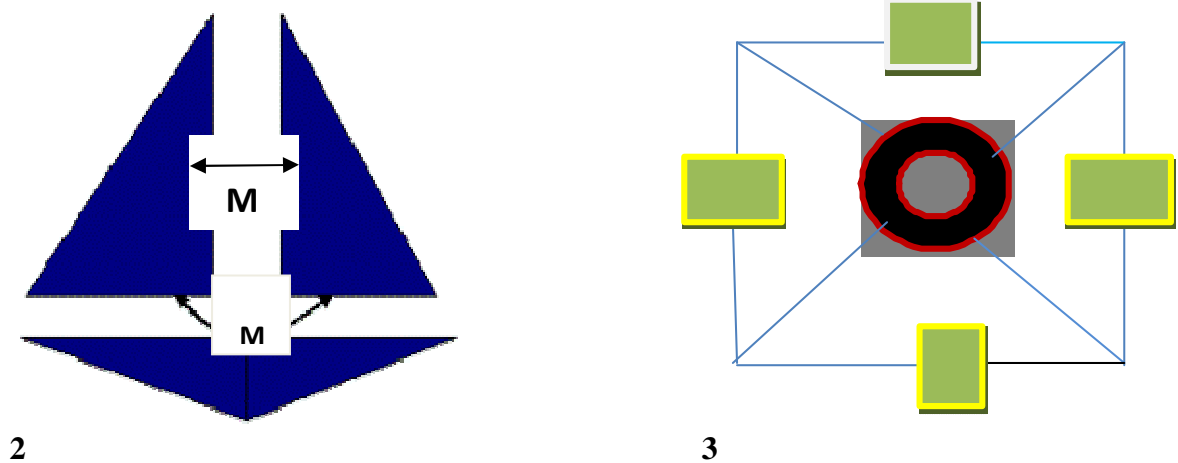

**2**   **3**

**Figure 2.** Three identical oscillators according to Figures 1A and 1B mutually coupled via M.

**Figure 3:** Schematic representation of four coupled electromagnetic circuits referring to eqs. (4.9 – 4.12). The inductivity L of each circuit is placed in the center and each oscillator is coupled to all other oscillators.

Using the canonical momentum of the three Lagrange functions we can either define the Hamilton operator (and Schrödinger equation) for $q_1$, $q_2$, $q_3$ or introduce creation- and annihilation operators as previously carried out. In the latter case, the Hamiltonian reads:

$H = \frac{\hbar}{2}[\omega_1 \cdot (b_1 b_1^+ + b_1^+ b_1) + \omega_2 \cdot (b_2 b_2^+ + b_2^+ b_2) + \omega_3 \cdot (b_3 b_3^+ + b_3^+ b_3)]$;

$[b_k, b_l^+] = \delta_{k,l}$ ($k,l = 1,..,3$)  (4.8)

It should be pointed out that the case of 3 independent oscillators (M = 0) implies the symmetry group $SU_3$ [9]. However, without any coupling a transfer of energy from one mode to another one cannot occur (the degeneracy is threefold). Due to the mutual coupling with $M \neq 0$ the symmetry is reduced to $SU_2$ and an additional hypercharge, i.e. $SU_2 \otimes U(1)$, denoted b $\lambda_3$ and $\omega_3$, since $\lambda_1 = \lambda_2$ and $\omega_1 = \omega_2$. In similar fashion, we can treat four mutually coupled oscillators with $M \neq 0$ (Figure 3):

$$L \cdot \ddot{Q}_1 + M \cdot (\ddot{Q}_2 + \ddot{Q}_3 + \ddot{Q}_4) + Q_1/C = 0;$$
$$L \cdot \ddot{Q}_2 + M \cdot (\ddot{Q}_1 + \ddot{Q}_3 + \ddot{Q}_4) + Q_2/C = 0;$$
$$L \cdot \ddot{Q}_3 + M \cdot (\ddot{Q}_1 + \ddot{Q}_2 + \ddot{Q}_4) + Q_3/C = 0;$$
$$L \cdot \ddot{Q}_4 + M \cdot (\ddot{Q}_1 + \ddot{Q}_2 + \ddot{Q}_3) + Q_4/C = 0 \quad (4.9)$$

Using similar substitutions to obtain the normal modes as above provides:
$$q_1 = Q_1 - Q_4;\ q_2 = Q_2 - Q_4;\ q_3 = Q_3 - Q_1;\ q_4 = Q_1 + Q_2 + Q_3 + Q_4 \quad (4.10)$$

Thus the normal modes are given by:
$$\lambda_1 = \lambda_2 = \lambda_3 = L - M;\ \lambda_4 = L + 3M;\ \omega_1^2 = \omega_2^2 = \omega_3^2 = \frac{1}{C(L-M)};\ \omega_4^2 = \frac{1}{C(L+3M)} \quad (4.11)$$

In terms of creation – and annihilation operators the Hamiltonian assumes the shape:

$$H = \tfrac{\hbar}{2} [\sum_{k=1}^{3} \omega_k \cdot (b_k b_k^+ + b_k^+ b_k) + \omega_4 \cdot (b_4 b_4^+ + b_4^+ b_4)];$$

$$[b_k, b_l^+] = \delta_{k,l}\ (k, l = 1, .., 4) \quad (4.12)$$

In particular, it has to be pointed out that $SU_3$ now is strictly conserved by the four mutually coupled circuits! The hypercharge now is related to the resonance oscillation $\omega_4^2 = \frac{1}{C(L+3M)}$. In principle, the formalism can be extended to higher order coupled circuits (e.g. 5$^{th}$ order). Then the normal modes are given by the relation:
$$\lambda_1 = \lambda_2 = \lambda_3 = \lambda_4 = L - M;\ \lambda_4 = L + 4M;\ \omega_1^2 = \omega_2^2 = \omega_3^2 = \omega_4^2 = \frac{1}{C(L-M)};\ \omega_5^2 = \frac{1}{C(L+4M)} \quad (4.13).$$

The Hamiltonian now is given by:
$$H = \tfrac{\hbar}{2} [\sum_{k=1}^{4} \omega_k \cdot (b_k b_k^+ + b_k^+ b_k) + \omega_5 \cdot (b_5 b_5^+ + b_5^+ b_5)];$$

$$[b_k, b_l^+] = \delta_{k,l}\ (k, l = 1, .., 5) \quad (4.14)$$

The physical content of the above equation incorporates a perturbed $SU_5$. Now the symmetry group $SU_4$ is exactly conserved. The remaining hypercharge formally increases due to the decreasing resonance frequency $\omega_5$. The inclusion of Ohm's resistance is also straightforward due to the elaborated modifications valid for the corresponding normal modes.

Finally we consider the case of a forced voltage $V_f$ in the circuit Figure 1A; the Schrödinger equation of this problem reads:

$$-\tfrac{\hbar^2}{2 \cdot L} \cdot exp(-R \cdot t / L) \cdot \tfrac{\partial^2}{\partial Q^2} \cdot \psi(Q,t) + [\tfrac{L \cdot \omega^2}{2} \cdot Q^2 - V_f \cdot Q \cdot exp(i \omega_f t)] \cdot$$
$$\cdot exp(R \cdot t / L) \cdot \psi(Q,t) = i \cdot \hbar \tfrac{\partial}{\partial t} \cdot \psi(Q,t) \quad (4.15)$$

Thus in many domains of applied physics damped circuits with forced oscillations are taken account for. The basis solution is the case $V_f = 0$. We consider this task based on time-dependent perturbation theory. The conventional perturbation expansions start with a time-independent Hamiltonian $H_0$ and a complete set of eigen-functions $\psi_n(x) \cdot exp(-i \cdot E_n \cdot t/\hbar)$. In our case the set of wave-functions (eq. (3.4)) is already time-dependent. The matrix elements of $H_I = V_f \cdot Q \cdot exp(i \cdot \omega_f \cdot t) \cdot exp(R \cdot t/L)$ result from the following integrations:

$$H_{I,nm}(t) = \int \psi_n(Q,t) \cdot V_f \cdot Q \cdot exp(i \cdot \omega_f \cdot t) \cdot exp(R \cdot t / L) \cdot \psi_m(Q,t) dQ \quad (4.16)$$

The expansion coefficient $c_m(t)$ is determined by:

$$i \cdot \hbar \tfrac{d}{dt} c_m(t) = \sum_{n=1}^{\infty} H_{I,nm}(t) \cdot c_n(t) \cdot exp(i(\omega(m-n)) \cdot t) \quad (4.17)$$

Due to the orthogonal properties of eq. (3.4) with regard to the variable $\xi$ we rewrite eq. (4.16) in the form:

$$H_{I,nm}(t) = V_f \cdot \sqrt{\tfrac{\hbar}{L \cdot \omega}} \cdot exp(-\tfrac{(m+n+1) \cdot R \cdot t}{2L}) \cdot \int_{-\infty}^{\infty} exp(-\xi^2) \cdot \xi \cdot H_m(\xi) \cdot H_n(\xi) d\xi \cdot exp(i \cdot \omega_f \cdot t) \quad (4.18)$$

The essential task is to evaluate the above integral via partial integration:

$$\int_{-\infty}^{\infty} exp(-\xi^2) \cdot \xi \cdot H_m(\xi) \cdot H_n(\xi) d\xi = -\tfrac{1}{2} \delta_{nm} \cdot n! \cdot 2^n \cdot \sqrt{\pi} + \tfrac{1}{2} \int_{-\infty}^{\infty} exp(-\xi^2) \cdot [H'_m(\xi) \cdot H_n(\xi) + H_m(\xi) \cdot H'_n(\xi)] d\xi \quad (4.19)$$

With the help of the relation $H_n'(\xi) = 2 \cdot n \cdot H_{n-1}(\xi)$ all terms provide:

$$\frac{1}{2}\int_{-\infty}^{\infty}exp(-\xi^2)\cdot[H'_m(\xi)\cdot H_n(\xi)+H_m(\xi)\cdot H'_n(\xi)]d\xi - \frac{1}{2}\delta_{nm}\cdot n!\cdot 2^n\cdot\sqrt{\pi} =$$

$$\frac{1}{2}2\cdot m\cdot[\delta_{m-1,n}\cdot 2^{m-1}\cdot(m-1)!\cdot\sqrt{\pi}] + \frac{1}{2}2\cdot n\cdot[\delta_{m,n-1}\cdot 2^{n-1}\cdot(n-1)!\cdot\sqrt{\pi}] - \frac{1}{2}\delta_{nm}\cdot n!\cdot 2^n\cdot\sqrt{\pi} \quad (4.20)$$

Evaluation of eq. (4.20) leads to the matrix elements of $H_I$:

$$H_{I,nm} = \sqrt{\pi}\cdot V_f\cdot\sqrt{\frac{\hbar}{L\cdot\omega}}\cdot exp(-\frac{(m+n+1)\cdot R\cdot t}{2L})\cdot exp(i\cdot\omega_f\cdot t)\cdot$$
$$\cdot\{m!\cdot 2^{m-1}{}_{(if:n=m-1)} + n!\cdot 2^{n-1}{}_{(if:m=n-1)} - n!\cdot 2^{n-1}{}_{(if:n=m)}\} \quad (4.21)$$

Inserting eq. (4.21) into eq. (4.17) provides:

$$i\cdot\hbar\frac{d}{dt}c_m(t) = \sqrt{\pi}\cdot V_f\cdot\sqrt{\frac{\hbar}{L\cdot\omega}}\cdot\sum_{n=0}^{\infty}c_n(t)\cdot exp(-\frac{(m+n+1)\cdot R\cdot t}{2L})\cdot$$
$$\cdot exp(i(\omega_f+\omega(m-n))\cdot t)\cdot\{m!\cdot 2^{m-1}\delta_{n,m-1} + n!\cdot 2^{n-1}\delta_{n,m+1} - n!\cdot 2^{n-1}\delta_{n,m}\}(4.22)$$

The first-order approach for the calculation of $c_{m,1}$ is obtained by the assumption that all terms of zero order are 1, then the integration provides:

$$c_{m,1} = \frac{1}{i\cdot\hbar}\cdot V_f\cdot\sqrt{\frac{\hbar\cdot\pi}{L\cdot\omega}}\cdot\sum_{n=0}^{\infty}exp(-\frac{(m+n+1)\cdot R\cdot t}{2\cdot L})\cdot[1 - \frac{2\cdot L}{(m+n+1)\cdot R}]\cdot$$
$$[exp(i\cdot(\omega_f+\omega(m-n))\cdot t)\cdot(\frac{1}{i\cdot(\omega_f+\omega(m-n))} - 1))]\cdot$$
$$\{m!\cdot 2^{m-1}\delta_{m-1,n} + n!\cdot 2^{n-1}\delta_{n-1,m} - n!\cdot 2^{n-1}\delta_{m,n}\} \quad (4.23)$$

This equation is usually solved iteratively by suitable boundary conditions, i.e. if t = 0 the exponential functions assume the value 1. The calculation of the second order approach is straightforward by inserting $c_{m,1}(t)$ into the right-hand side of eq. (4.22). Thus the most outstanding effect is that all terms of $c_{m,2}(t)$ are now proportional to the term $V_f^2$ according to eq. (4.24):

$$c_{m,2} = -V_f^2\cdot\frac{\pi}{\hbar\cdot L\cdot\omega}\cdot\sum_{n=0}^{\infty}exp(-\frac{(m+n+1)\cdot R\cdot t}{2\cdot L})\cdot[1 - \frac{2\cdot L}{(m+n+1)\cdot R}]^2\cdot$$
$$[exp(i\cdot(\omega_f+\omega(m-n))\cdot t)\cdot(\frac{1}{i\cdot(\omega_f+\omega(m-n))} - 1))]^2\cdot$$
$$\cdot\{m!\cdot 2^{m-1}\delta_{m-1,n} + n!\cdot 2^{n-1}\delta_{n-1,m} - n!\cdot 2^{n-1}\delta_{m,n}\} \quad (4.24)$$

In section 5 we shall consider two application cases with the forced oscillator in radiation physics. The resonance conditions of the quantum mechanical harmonic oscillator again can be verified from eq. (4.23) and eq. (4.24).

## 5. Some specific Applications

### 5.1 Excited States of $H_3PO_4$

With respect to the problem of cyclotron resonance in a constant magnetic field with friction in a medium considered at the end of section 2, we have already given an example with wide applications. Thus we can identify the charge $q_0$ of eq. (2.14 – 2.21) with an electron charge and friction with Ohm's resistance in a semiconductor, i.e. $D_{Friction} \approx R_{Ohm}$, then the reduced frequency provides the connection to the quantum Hall-effect. In section 3 the quantization of coupled electromagnetic circuits with damping has been presented. The coupled circuits implying the symmetry groups $SU_2$, $SU_3$, $SU_4$, etc. may have applications in nuclear physics and/or particle physics, if the parameters L, M and C are appropriately interpreted as distributions of baryonic charges and currents induced by mesons and gluons. In the domain of molecular physics, electric circuit oscillator models have first been introduced many years ago [10, 12], but only in the past decade they have obtained increasing importance. As an example we consider the excited states of phosphoric acid, which has already been calculated previously by two rather different methods, i.e. self-interacting field [10] and CNDOS-CI method [13]. However, the conformation of this acid depends on pH and solvent. Thus in a neutral medium (pH ≈ 7) this molecule is present by one double bond P = O (3d-electron), i.e. the summation formula is $H_3PO_4$. The configuration of this case is given by a $T_d$ structure. The lowest excited states of the perturbed $SU_4$ can readily associated with the discrete group $T_d$. The UV-absorptions starts at E = 4.8 eV. This fact also indicates that in this domain a double bond P = O is mainly involved in the excitation process. The remaining single P-OH bonds show absorption bands between 5.5 eV and 5.8 eV. Referring to the free $H_3PO_4$ (without interaction with a solvent) we have to account for the property that the double bond resulting from 3d electrons of phosphorus does not favor one of the four oxygen atoms. Thus from previous results [10] we have calculated $\hbar\omega_0$ = 5.31 eV and the energy shift due to the term MC. The calculated, perturbed $SU_4$ with regard to the four resonances amounts to 4.79 eV and 5.59 eV (3fold degeneracy). The interaction with the neutral solvent provides an excitation band between 4.76 eV and 4.84 eV and 5.52 eV and 5.77 eV. Therefore it is interesting to note that the damping influence, which is now mediated by dipole – dipole interactions of $H_3PO_4$ with the solvent and molecular vibrations, is accounted for the shifts given the corrected resonances:

$$\omega^2 \Rightarrow \omega^2 - R^2/4\lambda^2, \quad \lambda = L+3M \text{ or } \lambda = L-M \quad (5.1)$$

The resistance R realized by a solvent interaction and molecular vibrations can adopt the same function as the damping constant D in macroscopic physics.

### 5.2. Klystron

The quantum circuit model of a multi-cavity klystron can be treated according to Figure 1B, where the right-hand circuit additionally contains a forced oscillator $V_f \cdot \exp(i \cdot \omega_f t)$, but the left-hand circuit does not and emits waves with a much higher frequency. The condensers incorporate cylindrical cavities. The forced oscillator circuit (right-hand side) now is characterized by $\lambda_1 = L + M$; the left-hand condenser by $\lambda_2 = L - M$. Therefore, the inductivity L appearing in eqs. (4.21 – 4.23) have to be replaced by $\lambda_1 = L + M$. The action of the forced oscillator has a longer duration t = τ, i.e. τ >> $T_f$ and $T_f$ results from $\omega_f = 2\cdot\pi/T_f$. With regard to the mutual inductivity M it is useful that the solenoids of both circuits have a very narrow linkage, i.e. a chain-linking of the solenoids is realized. Assume: M = 0.9·L provides two resonance frequencies via $\lambda_1$ = 1.9·L and $\lambda_2$ = 0.1·L. The ratio between the both amounts to (if R is negligible):

$$\frac{\omega_2}{\omega_1} = 19 \quad (5.2)$$

An extremely strong coupling M = 0.99·L provides the ratio:

$$\frac{\omega_2}{\omega_1} = 199 \quad (5.3)$$

By that, the energy $Q \cdot V_f \cdot \exp(i \cdot \omega_f t)$ undergoes pumping to the second resonator, which can emit waves with a much higher frequency. If we add, from the left-hand side, a further oscillator circuit with a very low inductivity L (but M should assume values very close to L) and capacitance C, then we can reach a further drastic amplification of the frequency. I.e. the GHz domain will be reached, when e.g. via the relation $\omega_3/\omega_2$ = 200 the initial frequency of the forced oscillator was assumed to be $\nu_f$ = 200 Hz. The wave impedance $\rho_{I,k}$ of the oscillator k (k = 1,.., 3) is given by:

$$\rho_{I,k} = \frac{\lambda_k}{\sqrt{\lambda_k \cdot C - R^2 \cdot C^2/4}} \quad (5.4)$$

By that, the total impedance $\rho_I$ of the model klystron consisting of 3 coupled oscillators presented above has the form:

$$\rho_I = \sqrt{\sum_{k=1}^{3} \rho_{I,k}^2} \quad (5.5)$$

We should also point out that the relations (5.4) and (5.5) can certainly be optimized, when the LC-terms are taken different for each oscillator. This fact indicates that wave-guides can be designed by quantized, forced circuits, which are mutually coupled.

### 5.3. Bremsstrahlung

Typical atomic properties in terms of circuits have been presented in some publications [10 – 16], and further references can be found in the quotations. The creation of 'bremsstrahlung' is a particular quantum theoretical problem, which finds various important applications in rather wide fields, e.g. medicine [17], if the kinetic energy $E_{kinetic}$ of the impinging electron satisfies $E_{kinetic} \gg m_0 c^2$ (0.52 MeV). Therefore the 'bremsstrahlung' creation of electrons in metals ($E_{kinetic} \ll m_0 c^2$) being proportional to the nuclear charge Z is not the topic here. The theoretical base of this phenomenon has first been given by Bethe and Heitler [18]. However, the present study prefers the restriction given by viewpoint of Feynman [19], since we restrict ourselves here to principal method.

Figures A and B show an impinging electron moving in direction to the nucleus (A) and passing it with a deflection angle, which implies an increased virtual orbital. The connection of an LC-circuit to atomic properties have been previously been studied [14 – 16]. The motion of an electron, which represents a current, around the nucleus is connected to the inductivity $L$ and the charge distribution to the capacitance $C$. In similar fashion as previously carried out we have to fix the atomic parameters with regard to the virtual orbital according to Figure 4A. Since this presentation is preferably model consideration we assume that for the velocity **v** of electrons with $E_{kinetic} \gg 1$ MeV to be c (velocity of light). The creation of 'bremsstrahlung' of a single electron is a very fast process, but the behavior of the electron thereafter will be associated with a current and the resistance $R \neq 0$.

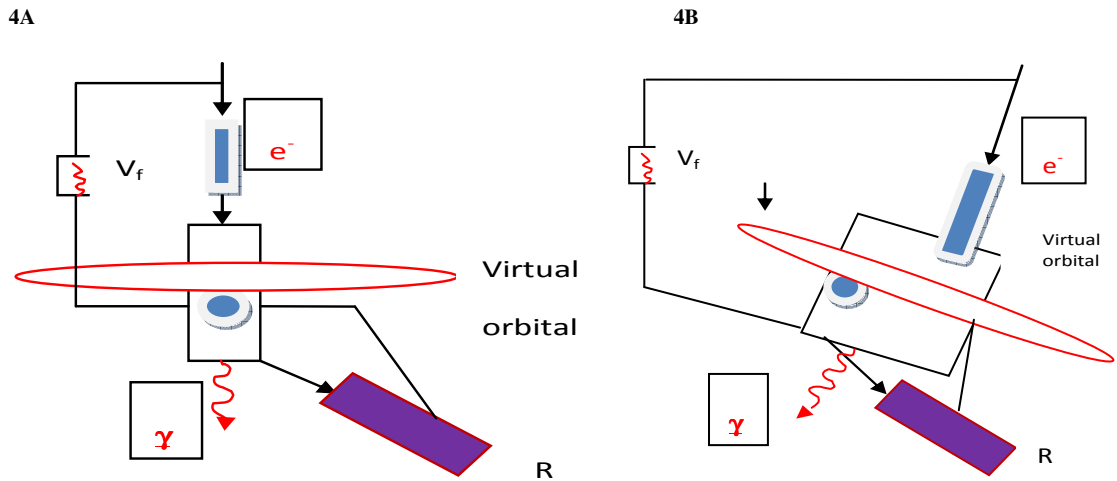

**Figure 4:** Circuits for the creation of 'bremsstrahlung' (smallest possible virtual orbital) – A. Impinging electron immediately hits the nucleus; B. Oblique impinging electron (the virtual orbital perpendicular to the impinging plane is increased). The energy loss due to the residual electron energy is denoted by the resistance R.

With the help of the propagator perturbation method [19] the photon creation with energy $\hbar \cdot \omega$ by an electron in the Coulomb field of a nucleus the differential cross-section reads:

$$\frac{d\sigma}{d\omega} = \frac{1}{2\pi} \cdot Z^2 e_0^6 \cdot \frac{1}{\omega} \cdot \frac{p_f}{p_i} \sin\theta_2 \sin\theta_1 d\theta_2 \, d\theta_1 \, d\phi \cdot F(E_i, E_f, p_i, p_f, \omega, \theta_1, \theta_2, \phi) \quad (5.6)$$

The function **F** detailed stated by Feynman [19] incorporates the dependence of the initial electron energy $E_i$, final energy $E_f$, and related momentums $p_i$, $p_f$. The angles $\theta_1, \theta_2$ (*electron*) and $\phi$ (*photon*) refer to the behavior after the central hit with the nucleus Z. Thus the relation (5.6) corresponds to Figure 4A with the general assumption that the outgoing electron and photon possess certain spherical angles different from 0. In Figure 4A we have already tacitly used that for $E_i \gg m_0 \cdot c^2$ (0.52 MeV) these angles are approximately 0, i.e. the created photon has the direction of the impinging electron. Figure 4B represents a more complex situation, which has been given by Bethe and Heitler [18].

For brevity we now consider the case 4A; the initial electron energy may be of the order between 6 MeV and 20 MeV, i.e. this order corresponds to linacs used in radiotherapy. The transition between virtual orbital state and ground state producing the bremsstrahlung photon $\hbar \cdot \omega$ is determined according to the perturbation equations (4.16 – 4.25) and the associated basic equation (4.15). In similar fashion as via eq. (6.1) and eq. (6.2) of the final section the impinging electron is characterized by its Compton wave length and forcing frequency:

$$\left.\begin{array}{l} e_0 \cdot V_{ex} = \hbar \cdot \omega_{ex} \\ \omega_f = \frac{2 \cdot \pi \cdot M \cdot c^2}{\hbar} \end{array}\right\} . \quad (5.7)$$

The calculation of the circuit associated to $L_{virtual}$ and $C_{virtual}$ of the virtual orbital follows elaborated standard methods, since the impinging electron incorporates a current and $C_{virtual}$ is associated with the Bohr radius of the hydrogen divided by Z (e.g. tungsten Z = 74). The energy of the electron in the final state is expressed by the resistance R. Thus we obtain in every order of perturbation theory the condition:

$$\left.\begin{array}{l} E_{photon} = \hbar \cdot \omega = \hbar \cdot \omega_0 - R^2/(4 \cdot L^2) \\ \hbar \cdot \omega_0 = \frac{1}{\sqrt{L_{virtual} \cdot C_{virtual}}} \end{array}\right\} . \quad (5.8)$$

The calculation of the transition probability, which proportional to the square of the result of eq. (4.24) or eq. (4.25), delivers the results of the Feynman method, if additional information like the direction of the momentums is introduced in connection to the impinging current. The usual restriction to electromagnetic circuits would ignore some space-depending properties such as current and charge density (**j**, ρ) and focus the complete problem to energy and time.

The case 4A and its generalization given by Figure 4B can be best treated by an extension of the Hamiltonian (4.15):

$$\frac{1}{2L} \cdot exp(-R \cdot t/L) \cdot ((\tfrac{\hbar}{i} \tfrac{\partial}{\partial Q} - f(t))^2 \cdot \psi(Q,t) + [\tfrac{L \cdot \omega^2}{2} \cdot Q^2 - V_f \cdot Q \cdot exp(i\omega_f t)] \cdot$$
$$\cdot exp(R \cdot t/L) \cdot \psi(Q,t) = i \cdot \hbar \tfrac{\partial}{\partial t} \cdot \psi(Q,t) \quad (5.9)$$

Thus function *f* incorporates an additional coupling to the magnetic flux, which provides the properties of the outgoing photon. The introduction of the function *f* is consistent with the canonical commutation relation P →Φ (magnetic flux) – *f*, which is expressed by [P, Q] = -i·ℏ. However, the classical equation of motion including a forcing voltage and damping now reads:

$$Ł = exp(R \cdot t/L) \cdot \{ \tfrac{L}{2} \cdot \dot{Q}^2 + L \cdot f(t) \cdot \dot{Q} + Q \cdot V_f(t) - \tfrac{1}{2C} \cdot Q^2 \} \quad (5.10)$$

This equation incorporates the absorption of energy by $V_f(t)$ due to the virtual circuit and stimulation of the emission of radiation ('bremsstrahlung'). Thus the case 4B increases the area of the capacitance C and, by that, the resonance frequency $\omega_0$ and related photon energy is decreased. With the help of eq. (5.10) the generalized Langevin equation reads:

$$L \cdot \ddot{Q} + L \cdot \dot{f}(t) + R \cdot \dot{Q} + \tfrac{1}{C} \cdot Q = V_f(t) \quad (5.11)$$

The 2$^{nd}$ term of eq. (5.11) describes a further source of loss, namely the diminishment of the magnetic flux independent of the collision of electrons with lattice vibrations.

### 6. Discussion and Outlook

We have considered the aspects of friction and resistance in the frame work of quantum mechanics. Thus friction may only have a quantum mechanical importance, if it can be restricted to typical atomic/molecular considerations such as dipole-dipole interactions with solvents or collisions with vibrations. Ohm's resistance principally appears in connection with lattice vibrations. The common feature is that in the microscopic domain irreversibility is introduced, i.e. energy is transferred to the environment. The quantization of circuits and its connection to self-interacting nonlinear fields with internal structure has at first been studied some decades ago [10]. However, in the meantime quantized electromagnetic circuits have been considered by many authors [14 – 16]. A main topic was the importance of these considerations in the rather novel field of quantum information processes. The inclusion of 'friction' in the form of typical influences of Ohm's resistance leading again to a reduced characteristic frequency incorporates the essential extension of the present study. Finally it should be pointed out that the equidistant harmonic oscillator levels of circuits may often reduce the applicability to molecular processes, which generally do not show these properties resulting from interacting many-body systems. The application of forced quantized circuits to problems, e.g. wave-guide due to a Klystron or creation of 'bremsstrahlung', indicates that the present study opens the door for problems of quantum electronics and even quantum electrodynamics.

The application of forced quantized circuits to problems like optimization of Klystrons indicates that the present study opens the door for problems of quantum electronics and even quantum electrodynamics. It should also pointed out that the mutual coupling M between three or four oscillators can readily extended to interaction problems of nuclear physics [8 – 10] with the help of introduction of meson currents between nucleons. The external force can be realized by the interaction of a charged particle with a nucleus. Let assume the charge of this external particle to be $q_0$, then the forcing perturbation oscillator assumes the form:

$$q_{ex} \cdot V_{ex} = \hbar \cdot \omega_{ex}. \quad (6.1)$$

The frequency $\omega_f$ appearing in the exponential function $\exp(i \cdot \omega_f \cdot t)$ of the forcing oscillator is different from $\omega_{ex}$ and results from the time-factor of plane waves, i.e.:

$$\omega_f = \frac{2 \cdot \pi \cdot M \cdot c^2}{\hbar}. \quad (6.2)$$

Thus the model of four coupled oscillators appears to be an attractive viewpoint, since due to the mutual interaction of these oscillators $SU_3$ is conserved and the total symmetry group is described by $SU_3$ x $U(1)$. The generators of $SU_3$ can be verified in a previous publication [10] and references therein. This final example makes evident that systems of quantized circuits may have applications in a rather field such nuclear excitations and reactions, but also with regard to quantum information problems.

**References**

1. Grabert, H.: *Projection Operator Techniques in Nonequilibrium Statistical Mechanics*. Springer Tracts in Modern Physics **95**. Berlin: Springer-Verlag. ISBN 3-540-11635-4 (1982).
2. Van Zon, R., Ciliberto, S., Cohen, E.: Power and Heat Fluctuation Theorems for Electric Circuits. Physical Review Letters **92**, 13 (2004): 130601.doi:10.1103/PhysRevLett.92.130601.
3. Friedrich, R., Peinke, J., Renner, Ch.: *How to Quantify Deterministic and Random Influences on the Statistics of the Foreign Exchange Market*, Phys. Rev. Lett. **84**, 5224 - 5227 (2000).
4. Schuch, D., Chung, K.H., Hartmann, H.: Non-Linear Schrödinger-type field equation for the description of dissipative Systems. 1. Derivation of the nonlinear field equation and one-dimensional example. J. Math. Phys. **24**, 1652-1660 (1983).
5. Schuch, D., Chung K.M., Hartmann, H.: Nonlinear Schrödinger-type field equation for the description of dissipative systems 3. Frictionally damped free motion as an example for an aperiodic motion, J. Math. Phys. **25**, 3086–3092 (1984).
6. Tsekov, R.: Nonlinear friction in quantum mechanics. Ann. Univ. Sofia, Faculty Physics **105,** 14-21 (2012) [arXiv 1003.0304].
7. Garashchuk, S., Dixit, V., Gu, B., Mazzuca, J.: The Schrödinger equation with friction from the quantum trajectory perspective. J. Chem. Phys. **7** 138, 5 (2013). DOI: 10.1063/1.4788832.
8. Ulmer, W., Matsinos, E.: Theoretical methods for the calculation of Bragg curves and 3D distributions of proton beams. Eur. Phys. J. Special Topics 190, 1–81 (2010) DOI: 10.1140/EJPST/e2010-01335-7
9. Ulmer, W.: The Role of Electron Capture and Energy Exchange of Positively Charged Particles Passing Through Matter Journal of Nuclear and Particle Physics **2** 77-86 (2012). DOI: 10.5923/j.jnpp.20120204.01
10. Ulmer, W.: On the representation of Atoms and Molecules as Self-interacting Field with Internal Structure. Theoretica Chimica Acta **55,** 179 – 205 (1980).
11. Ulmer, W., Cornélissen, G.: Coupled Electromagnetic Circuits and Their Connection to Quantum-Mechanical Resonance Interactions and Biorhythms. Open Journal of Biophysics, **3**, 253-274 (2013) http://dx.doi.org /10.4236/ojbiphy.2013.34031.
12. Hartmann, H., Stürmer, W.: Zur Darstellung molekularer Schwingungen durch mechanische und elektrische Oszillatoren. Z Naturforschung **6** a, 751 – 762 (1950).
13. Ulmer, W.: A Theoretical Study of $H_3PO_4$, nor-N-Mustard, and Cyclophosphamide. Z. Naturforsch. **34** c 658 – 669 (1979)
14. Louisell, W.H.: Quantum Statistical Properties of Radiation (Wiley, New York, 1973).
15. Yakymakha, O.L., Kalnibolotskij, Y.M.: Very-low-frequency resonance of MOSFET amplifier parameters. Solid-State Electronics, **38** 3, 661-671 (1995).
16. Devoret, M.H., Martinis, J.M.: Implementing Qubits with Superconducting Integrated Circuits. Quantum Information Processing **3** 1, 1 - 20 (2004).
17. Ulmer, W.: Creation of High Energy/Intensity Bremsstrahlung by a Multi-Target and Focusing of the Scattered Electrons by Small-Angle Backscatter at a Cone Wall and a Magnetic Field—Enhancement of the Outcome of Linear Accelerators in Radiotherapy. IJMPCERO **2** 147 – 160 (2013); http: //dx.doi.org/ 10.4236/ ijmpcero. 2013.24020.
18. Bethe, H.A., Heitler, W.: *On the stopping of fast particles and on the creation of positive electrons.* Proc. Phys. Soc. Lond. **146**, 83–112 (1934).
19. Feynman, R.P.: *Quantum Electrodynamics.* (W.A. Benjamin, New York, 1962).